# Holographic imaging with multimode, large free spectral range lasers in photorefractive sillenite crystals

E. A. Barbosa

Laboratório de Óptica Aplicada, Faculdade de Tecnologia de São Paulo,

CEETEPS - UNESP

Pça Cel Fernando Prestes, 30, São Paulo – SP, Brazil, CEP 01124 060

ebarbosa@fatecsp.br

**Abstract** - The holographic imaging of rigid objects with diode lasers emitting in many wavelengths in a sillenite $Bi_{12}TiO_{20}$ photorefractive crystal is both theoretically and experimentally investigated. It is shown that, due to the multi-wavelength emission and the typically large free spectral range of this light source, contour fringes appear on the holographic image corresponding to the surface relief, even in single-exposure recordings. The influence of the number of emitted modes on the fringe width is analysed, and the possible applications of the contour fringes in the field of optical metrology are pointed out.

Holographic recording with photorefractive media has been found many applications in the fields of fundamental and applied research [1]. Due to very important characteristics, like reversible recording with unlimited number of recording-erasure cycles, high image resolution and high storage capability, the photorefractive materials are currently of great interest in areas like information storage [2], pattern recognition [3], phase conjugation [4], optical image processing [5] and optical metrology [6,7].

Macroscopic parameters like response time, typical diffraction efficiency, response to wavelength, birefringence and optical activity distinguish the photorefractive materials from each other. These parameters determine the choice of the most suitable material for a specific application. Among the photorefractive crystals, the sillenite crystals ($Bi_{12}SiO_{20}$, $Bi_{12}TiO_{20}$ and $Bi_{12}GeO_{20}$) have interesting properties which make them very useful for image processing and for optical metrology, particularly for holographic interferometry [8]. Despite their typically lower diffraction efficiency, they have a much shorter response time, which is very desirable for the field of non-destructive testing. Moreover, the limitation of low diffraction efficiency can be overcome through the anisotropic diffraction properties of these crystals, allowing the cut-off of the transmitted wave and enabling the readout of the weak diffracted wave [9]. The $Bi_{12}TiO_{20}$ (BTO) crystals have the aditional advantage of a relatively low optical activity for the red light, which allows the highest possible diffraction efficiency for this wavelength range [10]. It is well-known that diode red lasers are among the most compact, available and low-cost commercial lasers, which permits the construction of robust, efficient, low-cost, BTO-based holographic interferometers. Many works studied the holographic recording with photorefractive materials (including also the amorphous polymers) by diode lasers [11-13]. However, the recording and imaging with red diode lasers and BTO crystals has not been extensively studied so far.

This work reports, for the first time to the author's knowledge, the appearance of stable interference fringes on holographic images of rigid, non-vibrating objects in single exposure recordings. A red (λ=670 nm) diode laser is employed as the light source and a BTO crystal is the holographic medium. In order to describe this effect a theoretical analysis based on the holographic recording of multiple volume holograms by a multimode laser in a sillenite crystal was developed. By this analysis, one can show that the fringes which cover the holographic image are the contour fringes of the studied object surface. The depth difference between two neighbouring fringes is determined by the wavelength interval between two adjacent laser modes, which, in its turn, is related to the laser free spectral range (FSR). It is shown that the contour fringes in single-exposure recordings are only observed when short-cavity, multimode lasers are employed.

1 Theoretical analysis

Consider two coherent, monochromatic waves of wavelength λ interfering in an optically active sillenite photorefractive crystal (PRC), as shown in figure 1. The resulting interference pattern is responsible for a spatial rearrangement of charge carriers inside the crystal by diffusion or drift. The resulting charge redistribution leads to a spatially non-uniform electric field which in its turn produces a refractive index grating of amplitude $\Delta n_r$ via the electrooptic effect. In the readout process, the reconstructed object wavefront is proportional to the hologram diffraction efficiency. If the sillenite crystal is cut in a [110] transverse electro-optic configuration, the diffraction efficiency η can be written as [14,15]

$$\eta = \left( \frac{n_0^3 r_{41} E_{sc}}{2\lambda \cos\alpha} \frac{\sin\rho L}{\rho L} \right)^2, \qquad (1)$$

where $n_0$ is the crystal refractive index, $r_{41}$ is the electro-optic coefficient, $2\alpha$ is the angle between the interfering beams, $\rho$ is the optical activity and $L$ is the crystal thickness. The electric field amplitude $E_{sc}$ generated by the redistributed space charges in the crystal is dependent on the modulation index $m$ of the interference pattern. In a purely diffusion recording regime, this relation is given by [14]

$$E_{sc} \cong imE_D \equiv iE_D \frac{2R^*S}{I_0}, \qquad (2)$$

where $R$ and $S$ are the amplitude of the reference and the signal (or object) beams respectively, $I_0$ is the total incident light intensity, and $E_D$ is the diffusion electric field. The superscript * denotes the complex conjugation of the wave amplitude. It is convenient to express $m$ in its complex form since this parameter carries the phase information about the interfering beams.

In this study the reference and the object beams are originated from a multimode laser. Considering the oscillation of $N$ modes, those beams can be expressed as

$$R_N(0) = R_0 \sum_{n=-(N-1)/2}^{n=(N-1)/2} A_n e^{i[(k+n\Delta k)\Gamma_R + \phi_n]}$$

$$S_N(0) = S_0 \sum_{n=-(N-1)/2}^{n=(N-1)/2} A_n e^{i[(k+n\Delta k)\Gamma_S + \phi_n]} \qquad (3)$$

where $\Delta k \,(= 2\pi\Delta\lambda/\lambda^2)$ is the wavenumber interval between two adjacent modes, $A_n$ is a real coefficient, and $\phi_n$ is the phase of the $n$-th mode at the laser output. The factors $\Gamma_S$ and

$\Gamma_R$ can be regarded as the optical paths of the object and the reference beams, respectively, and $k_S = k_R = 2\pi/\lambda \equiv k$.

Each mode (i.e., each wavelength) will contribute for the recording of its respective hologram and therefore the resulting grating will be a superposition of all holograms. Since different modes are not mutually coherent, the phase difference $\phi_n - \phi_m$ ($n \neq m$) is likely to vary randomly in time. Thus, the interference of different modes does not contribute to the holographic recording and therefore it does not affect the process of fringe formation on the holographic image. It is only responsible for a background light which lowers the grating amplitude $\Delta n_r$ and consequently decreases the overall diffraction efficiency of the resulting grating. Hence, it is convenient to express the electric field from equation (2) through the following matrix product:

$$E_{sc} \cong iE_D \frac{R_0 S_0}{I_0} e^{ik(\Gamma_S - \Gamma_R)} \sum_{n=-(N-1)/2}^{n=(N-1)/2} \sum_{m=-(N-1)/2}^{m=(N-1)/2} \delta_{n,m} A_n A_m e^{i\Delta k(n\Gamma_S - m\Gamma_R)}, \quad (4)$$

where Kronecker's delta $\delta_{n,m}$ was inserted in equation (4) for the reasons mentioned above. The diffraction efficiency is then given according to equations (1) and (4) by

$$\eta = \eta_0 \left[ \sum_{n=-(N-1)/2}^{n=(N-1)/2} A_n^2 e^{in\Delta k(\Gamma_S - \Gamma_R)} \right]^2, \quad (5)$$

where $\eta_0 \equiv \left( \frac{n_0^3 r_{41} E_D}{\lambda \cos\alpha} \frac{\sin\rho L}{\rho L} \frac{R_0 S_0}{I_0} \right)^2$.

In the readout by self-diffraction, the readout beam is the reference one. Such beam of intensity $I_R$ is diffracted by the hologram, producing the holographic reconstruction of the object beam whose intensity $I_S$ is obtained from equation (5):

$$I_S = \eta I_R = \eta_0 \left[ \sum_{n=-(N-1)/2}^{n=(N-1)/2} A_n^2 \, e^{in\Delta k(\Gamma_S - \Gamma_R)} \right]^2 I_R \qquad (6)$$

Consider the particular case for which $A_n=1$ in order to simplify the following analysis of surface profilometry by holographic recording with multimode lasers. Equation (6) can now be written as

$$I_S = \eta_0 \left\{ \frac{\sin[N\Delta k(\Gamma_S - \Gamma_R)/2]}{\sin[\Delta k(\Gamma_S - \Gamma_R)/2]} \right\}^2 I_R \qquad (7)$$

Figures 2a, 2b and 2c show the behaviour of $I_S$ from equation (7) as a function of $\Gamma_S$ for $\Gamma_R = 0$, $\pi/\Delta k$ and $2\pi/\Delta k$, respectively, with $\Delta k \approx 1.39$ rd/mm and $N = 4$. The intensity maxima from equation (7) shown in figure 2 correspond to the bright fringes on the reconstructed object image. A change in the value of $\Gamma_R$ leads to a phase shift and consequently the fringes run over the object surface, as it can be seen from figure 2. The analysis of such interference pattern allows the profilometry of the three-dimensional surface. According to equation (7), while the phase $\Delta k[\Gamma_{S,P} - \Gamma_R]$ of a given point $P$ on a bright fringe is

$$\Delta k(\Gamma_{S,P} - \Gamma_R) = 2\pi q, \qquad (8)$$

the light phase from a point $O$ laying on the next bright fringe is given by

$$\Delta k(\Gamma_{S,O} - \Gamma_R) = 2\pi(q+1), \qquad (9)$$

where $q = 1, 2, ...$ By combining the equations above one may easily determine the difference on the optical paths between points $O$ and $P$ (i.e., between any pair of adjacent fringes) as a function of the wavelength difference $\Delta\lambda$:

$$\Gamma_{S,O} - \Gamma_{S,P} = \frac{2\pi}{\Delta k} = \frac{\lambda^2}{\Delta\lambda} \qquad (10)$$

Notice by equation (10) that the difference $\Gamma_{S,O} - \Gamma_{S,P}$ is directly related to the laser FSR for longitudinal modes typically given by $\Delta\nu = c/2l = c\Delta\lambda/\lambda^2$, where $l$ is the laser resonator length. Hence, for long resonator ($l > 20$ cm) multimode lasers the holographic imaging of standard-size objects does not allow the identification of contour fringes, since the free-spectral range of such lasers is much smaller than that of short-length ($l \cong 1$ mm, typically) diode lasers.

Figures 3a, 3b and 3c show the intensity of the diffracted wave with respectively 2, 5 and 8 oscillating modes for $\Gamma_R = 0$, and the other parameters with the same values as in figure 2. For $N = 2$ the intensity has the expected $\cos^2$-type profile, as in a conventional

two-wavelength, double-exposure process. Notice that, the higher is the number of modes *N*, the narrower is the bright fringe. In fact, the width $\delta\Gamma_S$ of the bright fringe, defined as twice the $\Gamma_S$ difference from the fringe peak to the first zero, can be easily obtained with the help of equation (7) :

$$\delta\Gamma_S = \frac{2}{N\Delta k} \qquad (11)$$

**2 Experiments**

**2.1 Holographic recording with plane waves** – In order to verify the relation between the optical path of the interfering beams and the diffracted wave intensity given by equation (6), the holographic recording with plane waves was carried out. The optical setup is depicted in figure 4. The reference and the signal beams (*R* and *S* in figure 4, respectively) interfere at the BTO crystal after passing through the polarizer P1. Using the anisotropic diffraction properties of the sillenite crystals and considering their optical activity, if the input wave polarization is selected to be parallel to the [001] axis half-way through the crystal, the transmitted and the diffracted beams are orthogonally polarized at the crystal output. Thus, the transmitted reference wave is blocked by the polarizer P2 and only the diffracted signal beam is collected by the photodetector PD. The optical path of the signal beam is varied through the displacement of the 90º-prism PR, which is mounted on a micrometer. The reflection of the beam by the prism PR does not change the superposition of the interfering beams at the BTO crystal as the prism is displaced throughout the experiment.

Figure 5 shows the intensity of the diffracted signal beam as a function of the prism position $\Gamma_S$, with the value $\Gamma_R = 0$ set arbitrarily. The dotted curve is the measured diffracted wave intensity, while the solid curve is the fitting of the experimental data with the intensity given by equation (6), for $N = 4$ laser modes and $\Delta k \approx 1.18$ rd/mm. With the help of equation (10) one can determine the wavelength interval $\Delta\lambda$ between adjacent laser modes to be $\Delta\lambda = 0.082$ nm (for $\lambda=670$ nm), and the FSR to be $\approx 53.4$ GHz.

**2.2 Holographic imaging** - The holographic setup for the imaging of three-dimensional surfaces is shown in figure 6. The object is imaged onto the BTO crystal by the lens L2, while the reconstructed object wave is collected by the lens L3 to build the holographic object image at the CCD camera. By properly adjusting the polarizers P1 and P2 as described in the previous section, only the holographic image is displayed in a computer monitor for further processing. The mirror M3 is attached to a micrometer which introduces a phase shift in the reference beam.

The holographic image of a 50-mm long flat metallic bar, $30^o$-tilted with respect to the front face of the crystal (see figure 6), was recorded. Figure 7a shows the holographic image of the bar covered with the expected vertical and parallel contour fringes, while figure 7b shows the interferogram intensity profile along the line A-B from figure 7a. The x-axis is shown in figure 6. Notice that the appearance of narrow fringes in the intensity profile from figure 7b is in agreement with the results obtained in section 2.1, which show the oscillation of 4 laser modes.

By a simple measurement of the distance between bright (or dark) fringes, the depth difference $\Delta d$ along the y-axis (see fig. 6) between any pair of points laying on adjacent fringes along the line A-B (or lines parallel to it) was determined to be $\Delta d \approx 2.6$ mm. Considering that both the beam inciding on the object surface and the beam scattered from it propagate in nearly opposite directions one gets

$\Gamma_{S,O} - \Gamma_{S,P} \cong 2\Delta d \approx 5.2\ mm$. Again from equation (10) one gets $\Delta\lambda \approx 0.08\ nm$, which is in good agreement with the result for $\Delta\lambda$ obtained in the previous section.

As discussed in section 1, by phase-shifting the reference beam a three-dimensional scanning of the object can be carried out as the contour fringes run along its surface. This phase shift was accomplished by displacing the mirror M3 with a micrometer along the direction shown in figure 6. Figures 8a, 8b and 8c show the contour fringes on the holographic image of a 40-mm diameter, metallic cylinder, for $\Gamma_R = 0$, $\pi/\Delta k$ and $2\pi/\Delta k$, respectively. The position of mirror M3 for the value $\Gamma_R = 0$ was set arbitrarily. The change in the fringe position as the phase $\Delta k \Gamma_R / 2$ is varied can be clearly observed. As expected, a repetition of the interferogram occurred as the mirror M3 was displaced by multiple integers of ~ 2.6 mm.

**3 Conclusions**: This work reports the formation of interference fringes on the holographic image from rigid surfaces with $Bi_{12}TiO_{20}$ crystals when multimode, large FSR diode lasers are used as the light source. The theoretical analysis describes the recording of multiple volume gratings in the BTO crystal and obtains the diffraction efficiency of the resulting hologram.

Two relevant characteristics of the holographic imaging where shown both theoretically and experimentally: first, the diffraction efficiency can have values from zero up to a maximum, depending on the difference between the optical paths of the reference and the object beams, so that the holographic image of the three-dimensional object appears covered of contour fringes in a single-exposure holographic process; besides, the bright fringe width inversely depends on the number of the oscillating laser modes. The combination of these two properties does allow a very sharp, simple and both qualitative and quantitative evaluation of the relief of three-dimensional surfaces. The employ of laser emitting with more than two modes results in contour fringes which are narrower than that obtained in two-wavelength methods. Thus, the narrower the fringes, the higher the profilometry resolution. By phase-shifting the reference beam it is then possible to obtain a complete and accurate 3-D scanning of the object surface.

The results of this work establish also the criterium for the geometry of the optical setup in the case of holographic recording with both reference and signal plane waves. Since the diffraction efficiency is strongly dependent on the phase difference of both waves, and since this dependence becomes more critical as more modes oscillate, the optical path of the interfering beams must be carefully adjusted.

**Acknowledgements**: the author is grateful to Prof. Jaime Frejlich from Universidade Estadual de Campinas, for providing with the $Bi_{12}TiO_{20}$ crystal used in this work, to Prof. L. C. Albuquerque from Faculdade de Tecnologia de São Paulo and to M. Gesualdi from Universidade de São Paulo for fruitful discussions. This work was partially supported by the Fundação de Apoio à Tecnologia.

**References:**

1 – K. Buse: Appl. Phys. B **64**, 273 (1997);

2 – E. Chuang, D. Psaltis: Appl. Opt. **36**, 8445 (1997);

3 - M. S. Alam, J. Khoury: Proceedings of the SPIE Conference on Photorefractive Fiber and Crystal Devices: Materials, Optical Properties, and Applications, San Diego, California (2000);

4 - C.C. Chang, T.C. Chen, Hon-Fai Yau, P.X. Ye: Opt. Mat. **18**, 1 (2001);

5 – T.-C. Poon, P. P. Banerjee, *Contemporary Optical Image Processing With Matlab*, Vol.1 (Elsevier Science 2001);

6 - M.P. Georges, V.S. Scauflaire and P.C. Lemaire: Appl. Phys. B: Lasers and Optics **72**, 761 (2001);

7 – J. Frejlich, P. M. Garcia: Optics and Lasers in Engineering **32**, 515 (2000);

8 – E.A. Barbosa, M. Muramatsu: Opt. Laser Technology **29**, 359 (1997);

9 – S. Mallick, D. Rouède: Appl. Phys. B: Lasers and Optics **43**, 239 (1987);

10 – P. Günter, J.-P. Huignard (eds.): *Photorefractive Materials and Their Applications II*, Topics Appl. Phys., Vol. 62 (Springer Berlin Heidelberg 1989);

11 - J.E. Millerd, N.J. Brock: Appl. Opt. 36, 2427 (1997);

12 – C. Yang, S. Yokoyama, T. Honda, K. Seta: Appl. Phys. Lett.**72**, 633 (1998 );

13 – S. MacCormack, R. W. Eason: Opt. Lett. **16**, 705 (1991);

14 – S. V. Miridonov, A.A. Kamshilin, E. Barbosa: J. Opt. Soc. Am. A **11**, 1780 (1994);

15 - E.A. Barbosa, J. Frejlich, V.V. Prokofief, N. J. H.Gallo, J. P.Andreeta, Opt. Eng. 33, 2659 (1994).

**Figure Captions**

Figure 1: Incidence of the reference (R) and the signal (S) waves onto the $Bi_{12}TiO_{20}$ crystal.

Figure 2: intensity of the diffracted beam as a function of $\Gamma_S$ for $N = 4$ and for a) $\Gamma_R = 0$; b) $\Gamma_R = \pi/\Delta k$ and c) $\Gamma_R = 2\pi/\Delta k$.

Figure 3: intensity profile of the interferogram as a function of $\Gamma_S$ for a) $N = 2$; b) $N = 5$ and c) $N = 8$ oscillating modes.

Figure 4: Optical setup for holographic recording with plane waves. BTO, $Bi_{12}TiO_{20}$ crystal; BS, beam sppliter; M1 and M2, mirrors; PR, $90°$-prism; P1 and P2, polarizers; PD, photodetector.

Figure 5: Diffracted beam intensity as a function of the signal beam optical path; dots, experimental data; solid line, fitting for $\Delta k = 1.18$ rd/mm and $N = 4$ laser modes.

Figure 6: Optical setup for holographic imaging; L1, L2 and L3, lenses; CCD camera and PC computer for image analysis.

Figure 7: a) contour fringes of the $30°$-tilted bar; b) light intensity profile of the interferogram along the line A-B.

Figure 8: Holographic image of the 40-mm diameter cilinder for a) $\Gamma_R = 0$; b) $\Gamma_R = \pi/\Delta k$ and c) $\Gamma_R = 2\pi/\Delta k$.

Figures

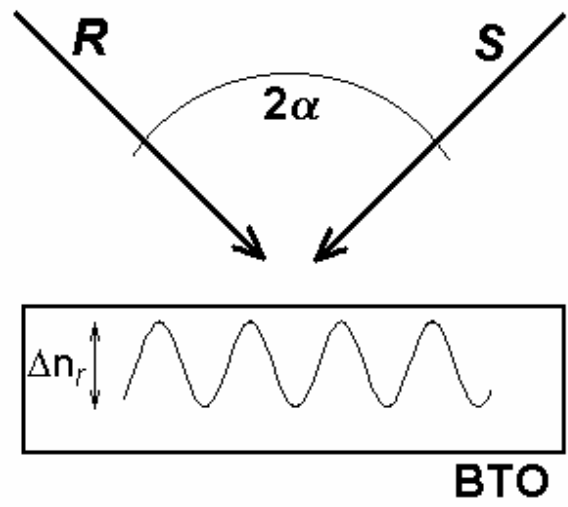

Figure 1

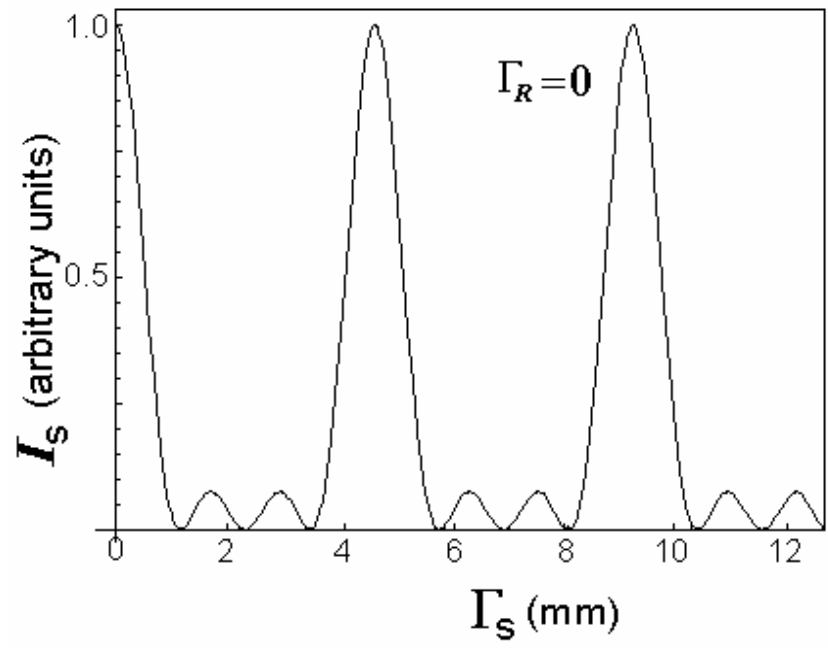

Figure 2a

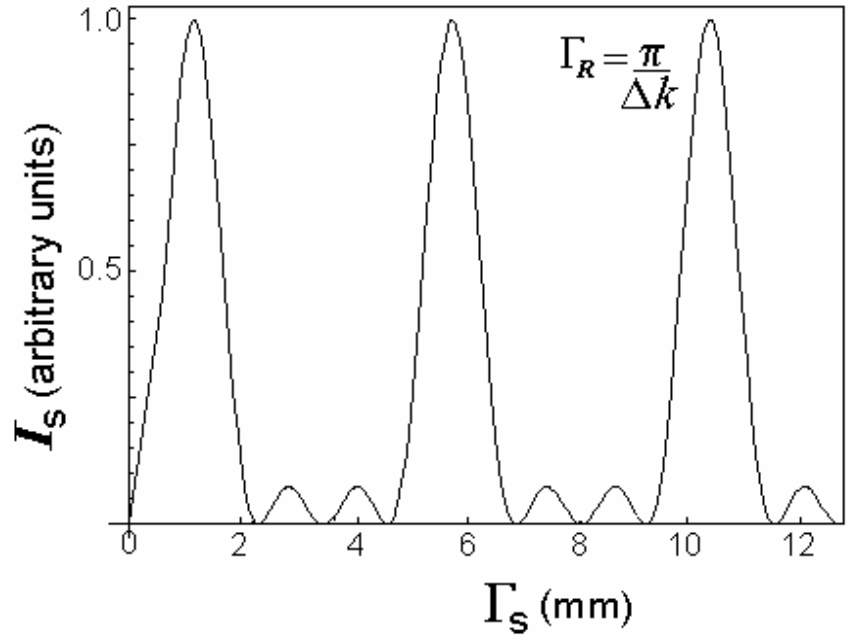

Figure 2b

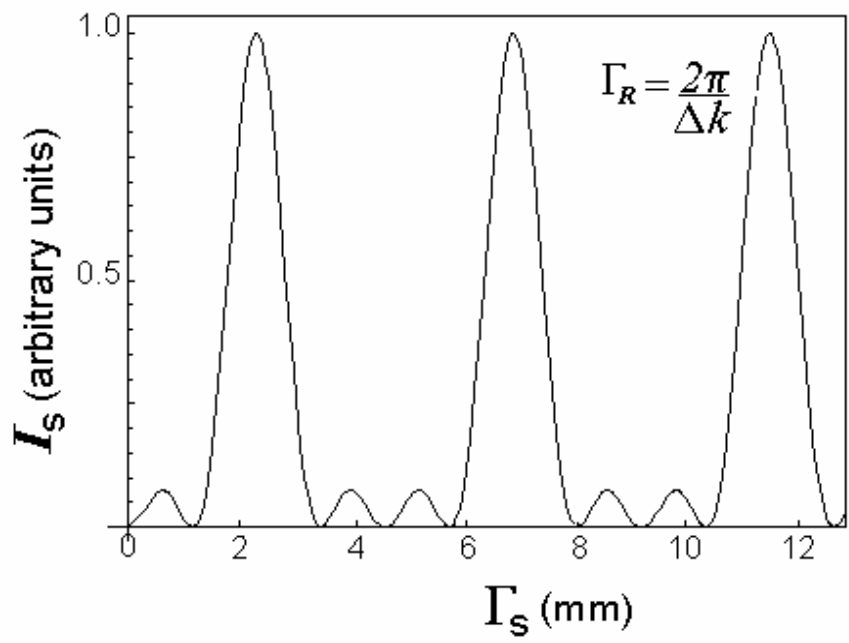

Figure 2c

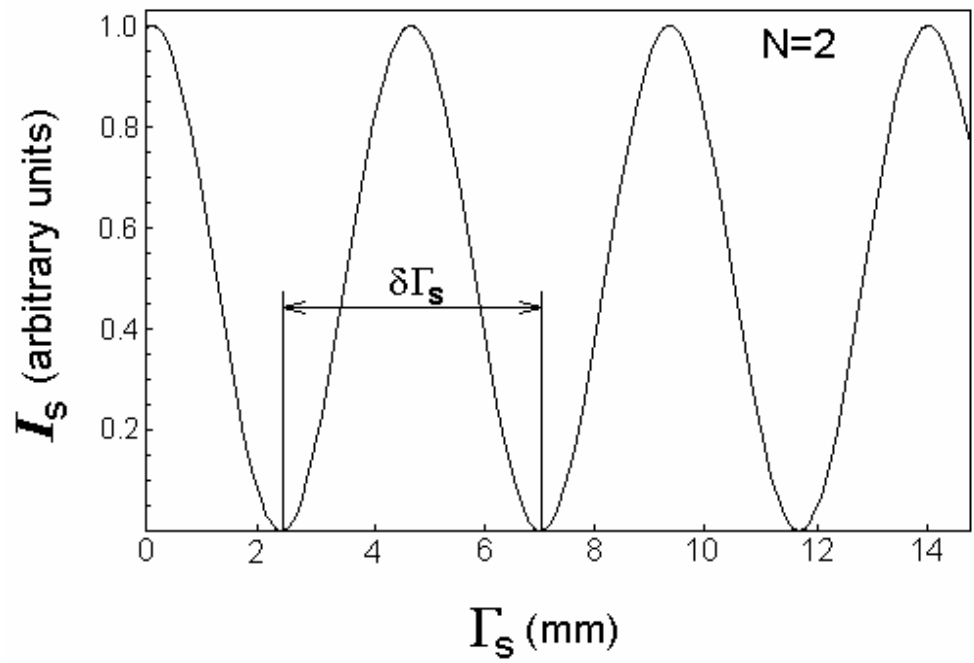

Figure 3a

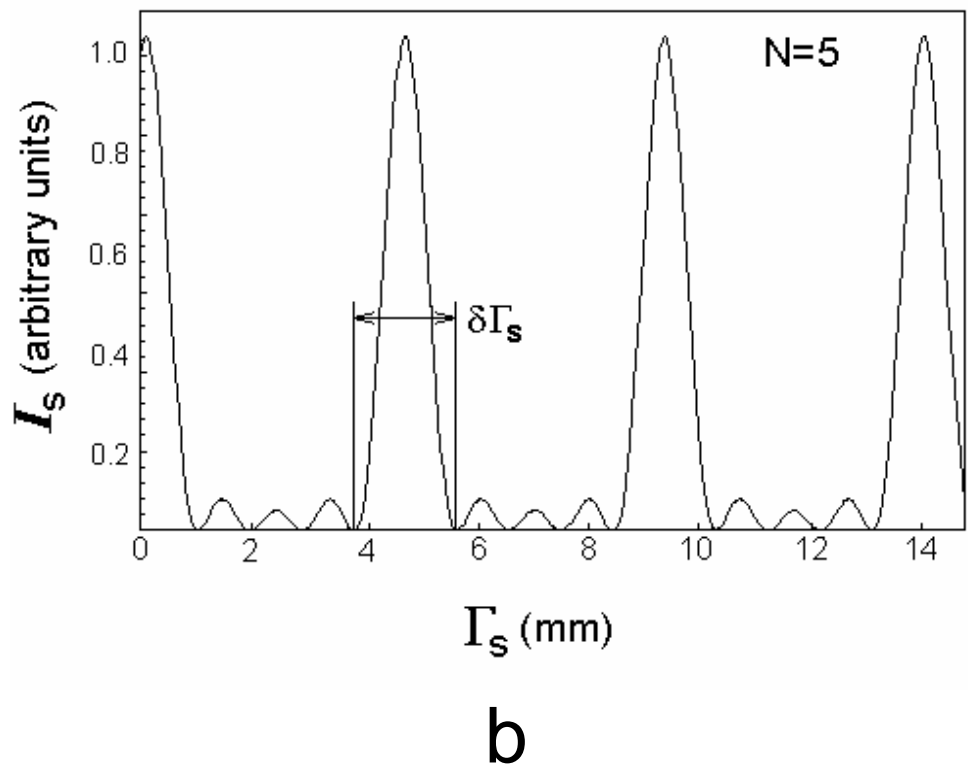

Figure 3b

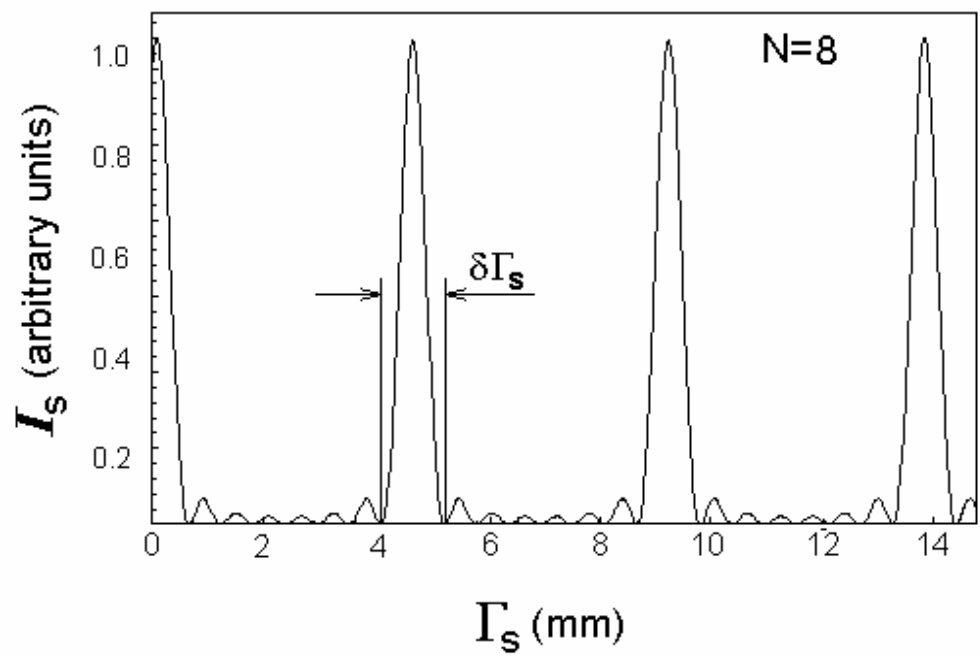

Figure 3c

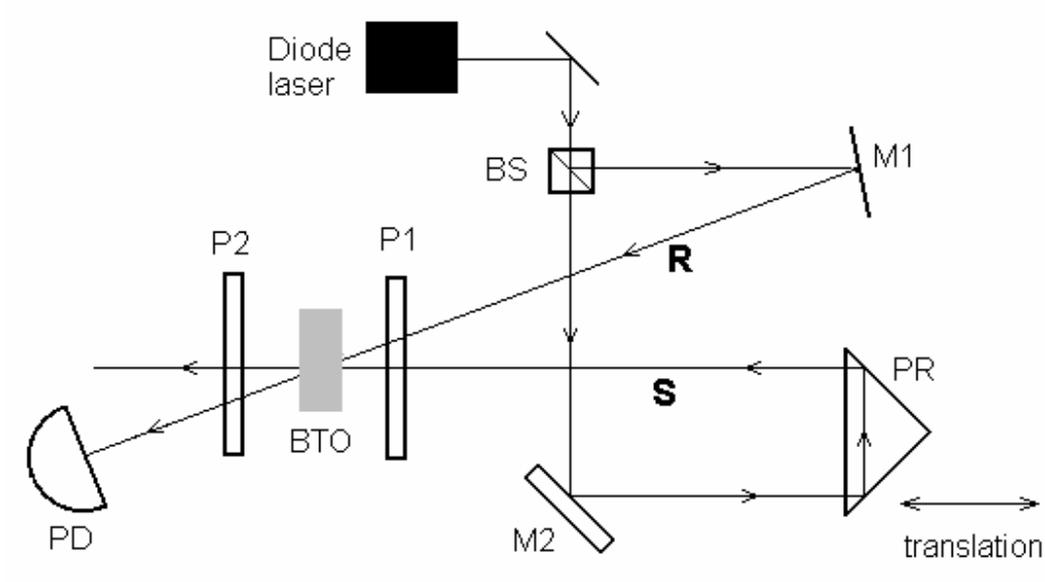

Figure 4

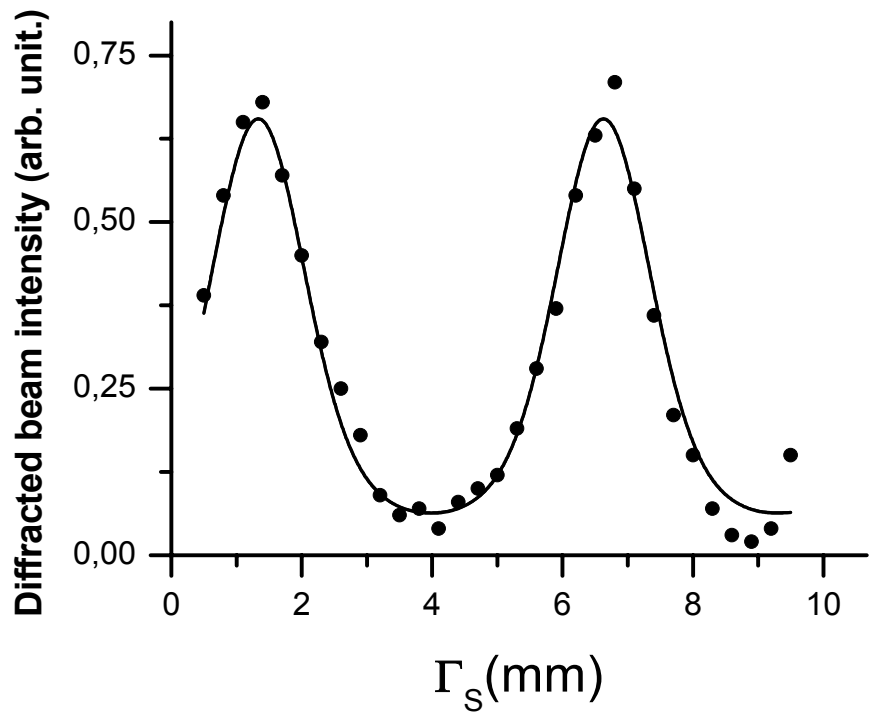

Figure 5

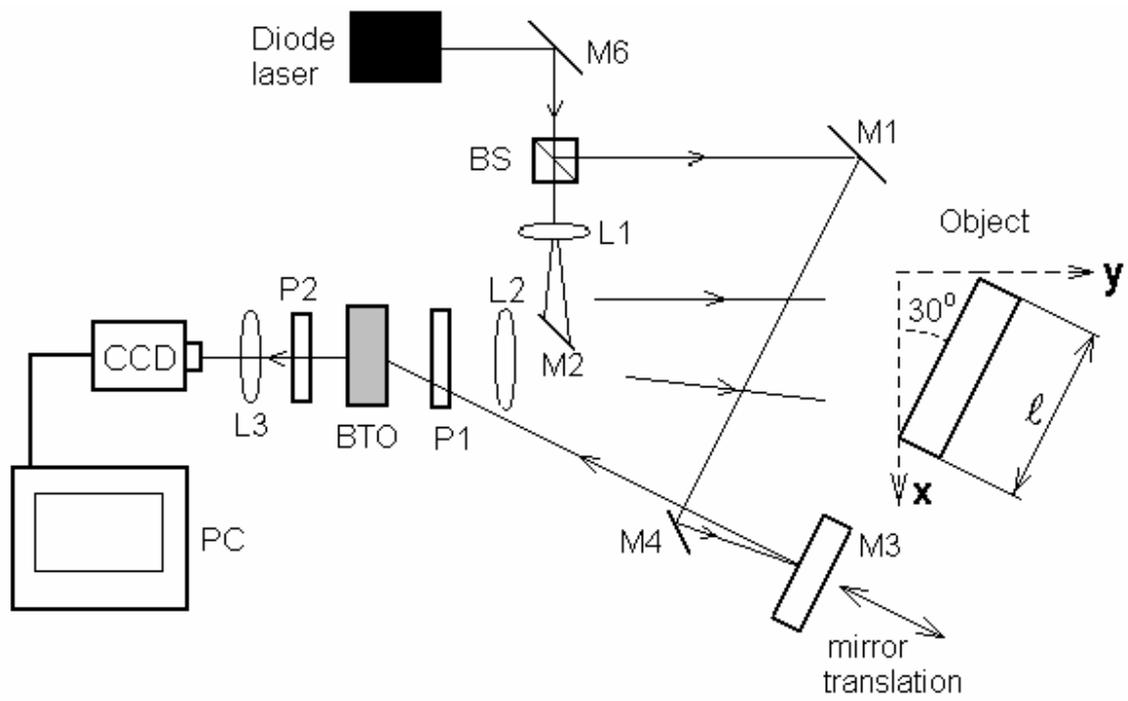

Figure 6

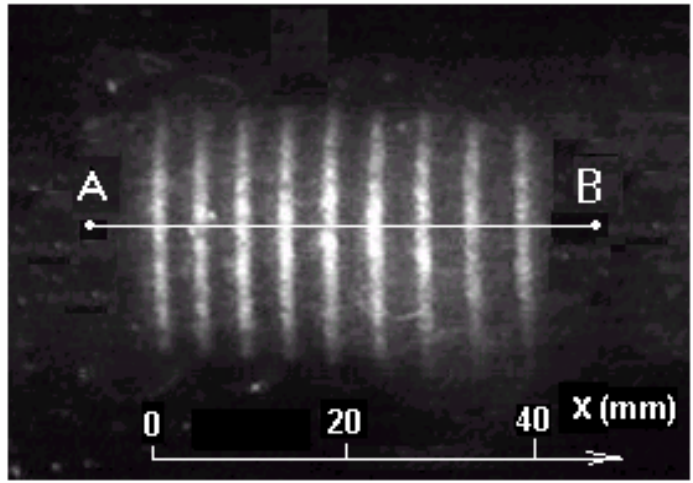

Figure 7a

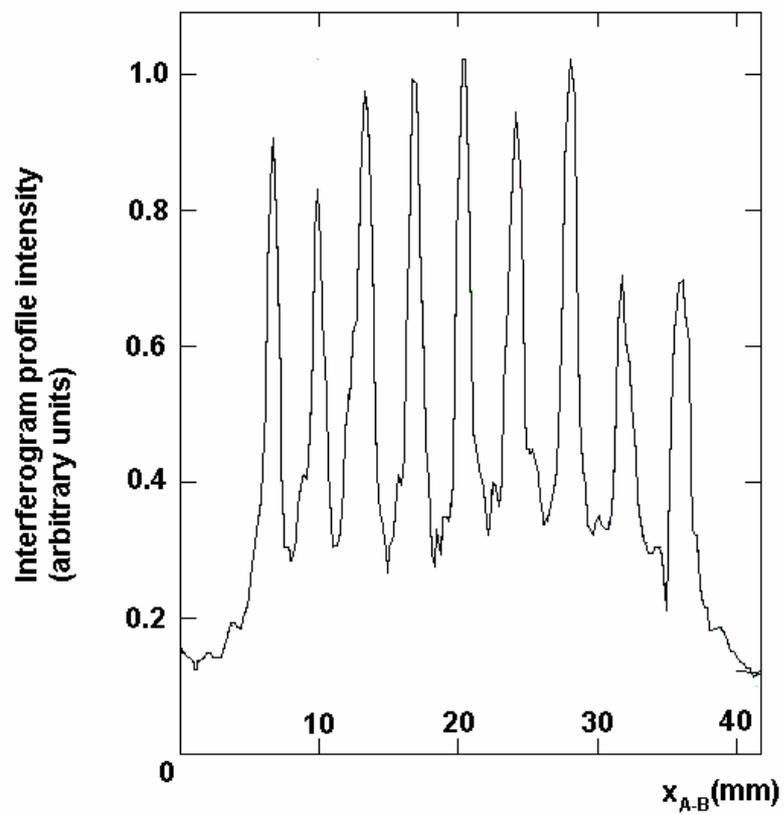

Figure 7b

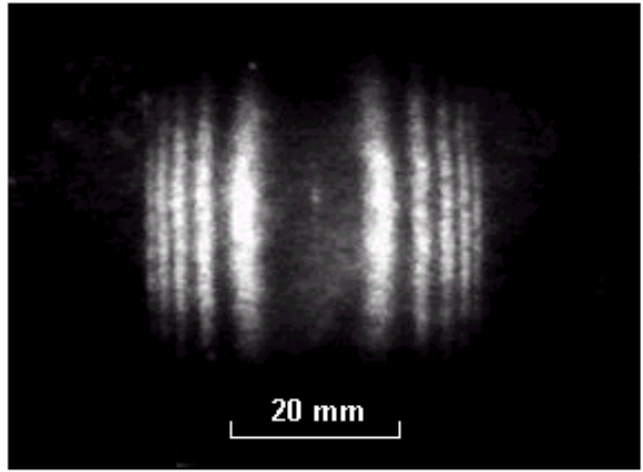

a

Figure 8a

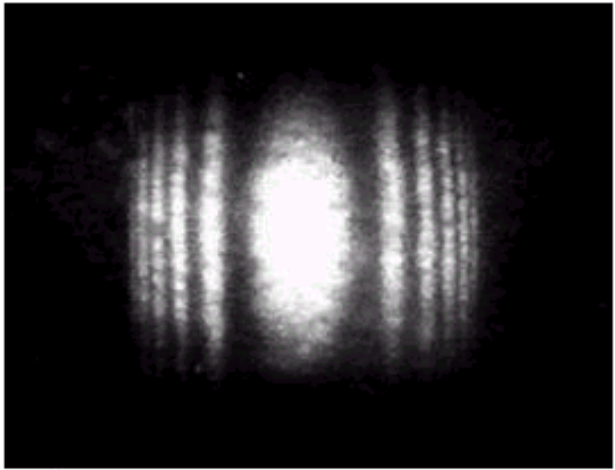

b

Figure 8b

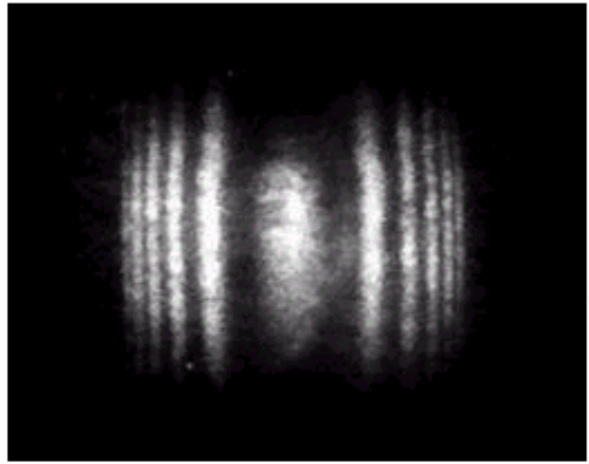

c

Figure 8c